\documentclass[bibyear]{aa}

\usepackage{natbib}

\usepackage{amsmath}	
\usepackage{gensymb} 	
\usepackage{orcidlink}
\usepackage{float}
\usepackage{comment}
\usepackage{hyperref}
\usepackage[varg]{txfonts}
\def\um{\mu \rm{m}}    
\def\Tb{t_{\rm b}}     
\def\tcool{t_{\rm cool}} 
\def\incmut{\theta_{\rm mut}}  
\newcommand\vhat[1]{\overset{\rightharpoonup}{#1}}

\begin{document}

\title{Brightness variability in polar circumbinary disks}

\author
{
I. Rabago \orcidlink{0000-0001-5008-2794} \inst{1,2} \and
G. Lodato \orcidlink{0000-0002-2357-7692} \inst{1} \and
S. Facchini \orcidlink{0000-0003-4689-2684} \inst{1} \and
Z. Zhu \orcidlink{0000-0003-3616-6822} \inst{2}
}
\institute{Dipartimento di Fisica, Università degli Studi di Milano, Via Celoria 16, 20133 Milano, Italy\\
\and
Department of Physics and Astronomy, University of Nevada, Las Vegas 4505 S. Maryland Parkway Las Vegas, NV 89154, USA\\
\email{ian.rabago@unimi.it}
}

\date{Received XXX; Accepted YYY}

\abstract{In binary systems with a strongly misaligned disk, the central binary stars can travel a significant vertical distance above and below the disk's orbital plane.  This can cause large changes in illumination of the disk over the course of the binary orbital period.  We use both analytic and radiative transfer models to examine the effect of changes in stellar illumination on the appearance of the disk, particularly in the case of the polar disk HD 98800B.  We find that the observed flux from the disk can vary significantly over the binary orbital period, producing a periodically varying lightcurve which peaks twice each binary orbit.  The amount of flux variation is strongly influenced by the disk geometry.  We suggest that these flux variations produce several observable signatures, and that these observables may provide constraints on different properties of the disk such as its vertical structure, geometry, and cooling rate.}


\maketitle



\section{Introduction}
In recent years, the number of protoplanetary disks observed in multi-star systems has increased dramatically.  These systems include disks in binary systems (HD 142527; \citealt{Avenhaus2014,Avenhaus2017}), triple systems (GW Orionis; \citealt{Kraus2020}), and systems of even higher multiplicity (GG Tau; \citealt{DiFolco2014,Keppler2020}).  The presence of more stars in these systems increases the complexity of the star-disk dynamics, generating new behaviors in the disk and providing additional probes into the nature of protoplanetary disks when combined with single-star observations.

For a handful of these systems, the orientation of both the disk and stellar orbital planes can be measured using a combination of mm-wavelength and astrometric observations, allowing for measurement of the misalignment between the two planes.  The observed misalignment angles are mostly coplanar for small binaries, but become increasingly misaligned for wider or more eccentric binaries \citep{Czekala2019}, including disks in ``polar'' orbits, with a misalignment angle close to $90^\circ$.  In a circumbinary disk, such a configuration can naturally arise and remain in a stable configuration as long as the central binary is eccentric \citep{Verrier2009,Farago2010}.  The evolution towards polar alignment in a highly inclined circumbinary disk has been studied carefully in many previous works \citep{Aly2015,Martin2017,Martin2018,Rabago2023}.  For higher order systems, the evolution of the disk is more complicated due to additional effects from the interacting bodies, but analytical results seem to suggest a certain amount of misalignment in these systems \citep{Ceppi2023,Lepp2023}.

Disks such as 99 Herculis \citep{Kennedy2012}, HD 98800B \citep{Kennedy2019}, and Bernahrd-2 \citep{Hu2024} are thought to be examples of disks in a polar orientation.  The HD 98800 quadruple system, consisting of a double binary pair \citep{Walker1988}, is the clearest example of a polar-aligned disk due to recent ALMA observations directly resolving the circumbinary disk around the B binary \citep{Kennedy2019}.  The observed orientation of the disk is consistent with a configuration that is only a few degrees away from exactly polar.

Polar disks offer a unique configuration in which the high misalignment of the binary and disk planes allows the central stars to gain significant vertical distances above and below the disk as the stars orbit.  This added vertical motion can greatly change the amount of light received across the face of the disk, resulting in significant changes in the appearance of the disk over the course of the binary orbital period.  This suggests an innate source of variability in systems with a polar-aligned disk, originating from periodic motion of the central binary.

In this paper, we examine how the movement of the central binary changes the illumination of the surrounding circumbinary disk.  We focus on the case of a polar disk to consider the case of maximum misalignment between the binary and disk planes.  The paper is organized as follows.  In Section \ref{sec:lightcurves} we examine the analytical variations in stellar flux received by the disk over the course of the binary orbital period.  In Section \ref{sec:radmc} we perform radiative transfer calculations for a disk in a polar configuration.  We discuss our results in Section \ref{sec:discussion} and conclude in Section \ref{sec:conclusions}.

\section{Numerical Light Curves}
\label{sec:lightcurves}
In this section we derive a numerical model for the amount of illumination received by the disk.  We consider the arrangement of an equal-mass point sources (the binary) with separation $a$ and eccentricity $e$ and a surrounding polar-aligned annulus of material (the disk).  Polar alignment occurs when the disk angular momentum vector aligns with the binary eccentricity vector.  In the following discussion, we place the binary-disk system with their geometric center at the origin, and orient the system such that the disk lies along the $xy$-plane and the binary major axis and eccentricity vector lie in the $z$ direction.  Without loss of generality, we take the plane of the binary orbit to be in the $xz$-plane.

We consider a disk with a radial extent of $2a$ to $5a$, similar to the polar disk of HD 98800B \citep{Kennedy2019}.  The inner radial limit is determined by the truncation of the central binary, and is slightly smaller than the expected inner cavity for coplanar circumbinary disks \citep{Miranda2015,Franchini2019}; the outer radius of HD 98800B sits at around $5a$, and is likely truncated by the orbit of the AaAb binary \citep{Zuniga-Fernandez2021}.  When calculating the stellar flux onto the disk, we consider the stars to be point-like sources.

The incoming stellar flux across the entire disk is given by:

\begin{equation}
    F_{\rm tot} = \iint \frac{F_* \cos{\theta} (r,\phi, r_*(t))}{d(r,\phi,r_*(t))^2} \ rdrd\phi,
    \label{eq:diskflux}
\end{equation}
where $F_*$ is the flux from the stellar surface, $d = ||\vhat{\mathbf{r}}-\vhat{\mathbf{r}}_*||$ is the distance from the star to a point on the disk, and $\theta$ is the angle between the star and the normal of the disk surface in question ($\tau = 1$ surface, scattering surface, etc.).  In this equation, $d$ and $\theta$ are functions of both the location on the disk and the position of the illuminating star, as well as a function of time as the star moves along its orbit.  In the following sections, we discuss some specific disk models and the important features of each.

\subsection{Thin disk}
We start with a simple model of a razor-thin disk, $h/r \ll 1$.  In this case, the disk surface is coincident with the $xy$-plane and the disk normal points vertically in the $z$ direction.  At large distances from the binary, the lateral motion of the stars in the direction of the disk plane is negligible and we can assume the star to lie along the $z$-axis, so the flux contribution at each point takes the form:

\begin{equation}
    F = \frac{F_*}{d^2} \cos\theta = F_* \left(\frac{z}{d^3}\right),
\end{equation}
where $z$ is the vertical distance of the star above the disk plane.  Thus, similar to a dipole, the incoming flux should decay as the cube of the radial distance.  Inner regions of the disk are expected to receive substantially more radiation, as the stellar flux can be angled more directly onto the disk surface during binary apocenter.  As binary eccentricity increases, both the apocenter distance $a(1+e)$ and the fraction of time the stars spend close to apocenter increase, so we expect the effects of face-on radiation to increase for higher eccentricity binaries.

\begin{figure}
    \centering
    \includegraphics[width=\columnwidth]{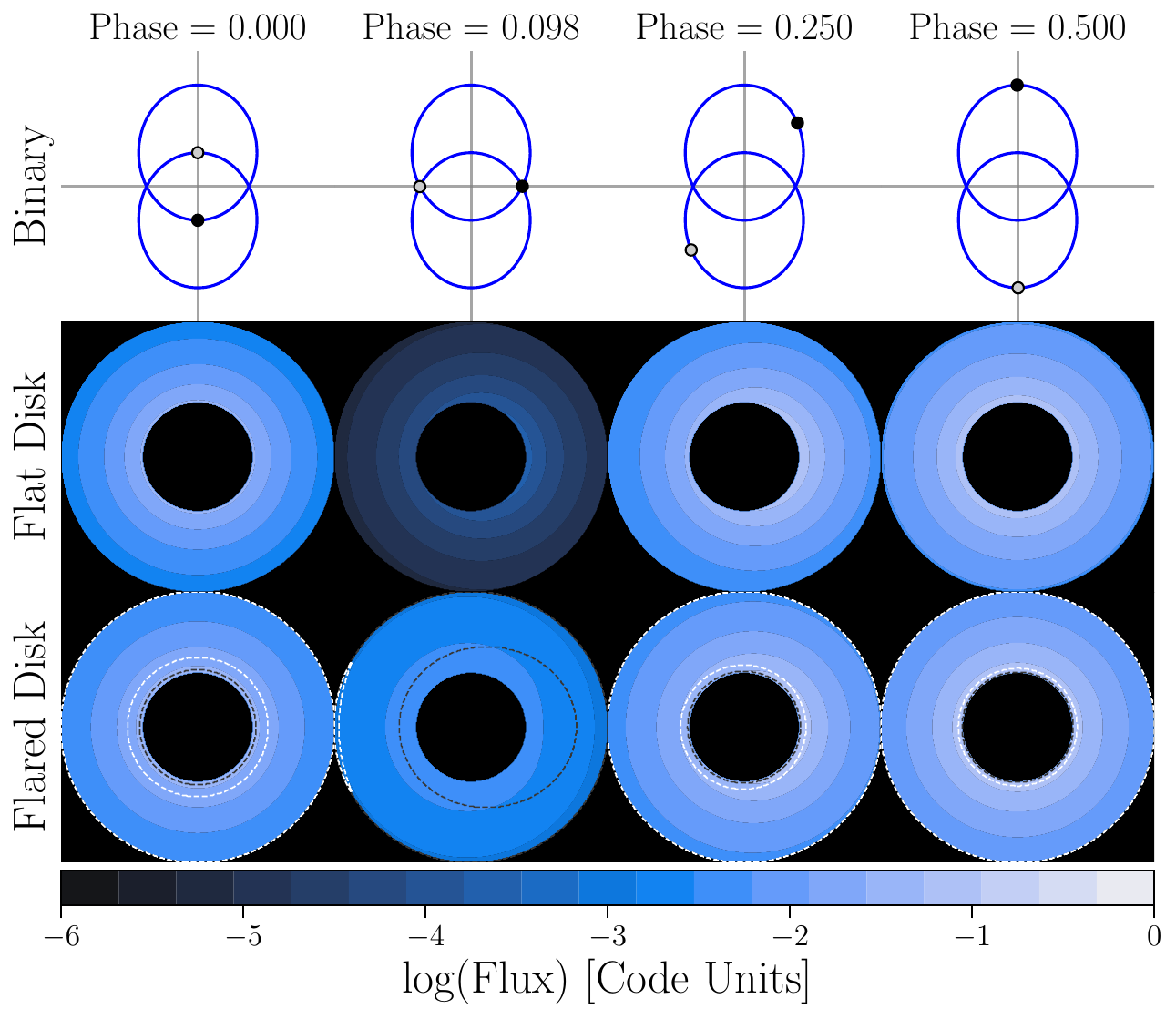}
    \caption{Illumination of the disk at different points of the binary orbit, for an equal mass central binary of $e=0.5$.  Each column shows the position of the binary stars (top), along with the illuminated flux received by a razor-thin disk (middle) and a flared disk (bottom).  For the flared disk model, dashed lines along the disk face indicate the regions where the incoming starglight would be shadowed due to the presence of a puffed inner rim; gray and white dashed lines show shadowed regions for puff heights of $1.5h_{\rm in}$ and $2.0h_{\rm in}$, respectively.  Note that, for the second column, the stars are shown with a small vertical displacement to show asymmetric lighting effects close to plane crossing. }
    \label{fig:diskillumination}
\end{figure}

To examine how the incoming stellar flux changes as the stars move, we numerically calculate the flux across the disk over the binary orbital period.  Using Kepler's equation, we solve for the positions of the binary components at 1000 evenly spaced points in time (i.e. evenly spaced in the mean anomaly $M$) and use Equation \ref{eq:diskflux} to calculate the total flux onto the disk at each time.  Following the convention used for orbital astrometry, we normalize the orbital period of the binary from 0 to 1, with $t=0$ representing the time of pericenter passage for the binary system.  Note that, for these calculations, we consider the disk to be opaque, and consider the light of only one star at a time for one side of the disk.  Depending on the optical depth of the disk, radiation from both stars may contribute to the total illumination in a more realistic scenario.  For an equal-mass binary with stars of roughly the same temperature, the results should change by a factor of two at most.

Figure \ref{fig:diskillumination} shows the flux onto the disk at different points of the binary orbit, for an equal-mass central binary with eccentricity $e=0.5$.  Notably, the minimum brightness does not occur at binary pericenter as one might expect, as although the stars are at their closest they are both vertically offset from their center of mass, allowing them to illuminate the surface of the disk a small amount (first column).  Instead, the minimum illumination occurs slightly before and after pericenter (second column), when the binary stars pass through the disk plane.  At these times, which we refer to hereafter as disk plane crossings, the stellar flux runs parallel to the disk and, in our razor-thin approximation, the illumination drops to zero.  As the stars continue in the orbit (third column), their vertical and horizontal distance from the system center increases, and the disk brightens considerably as the star now shines down onto the face of the disk directly.  The lateral displacement of the stars at mid-orbit creates an asymmetric illumination pattern along the axis of the binary plane.  Finally, as the stars reach apocenter the horizontal displacement vanishes, leaving a symmetric, brightly illuminated disk (fourth column).

\begin{figure}
    \centering
    \includegraphics[width=\columnwidth]{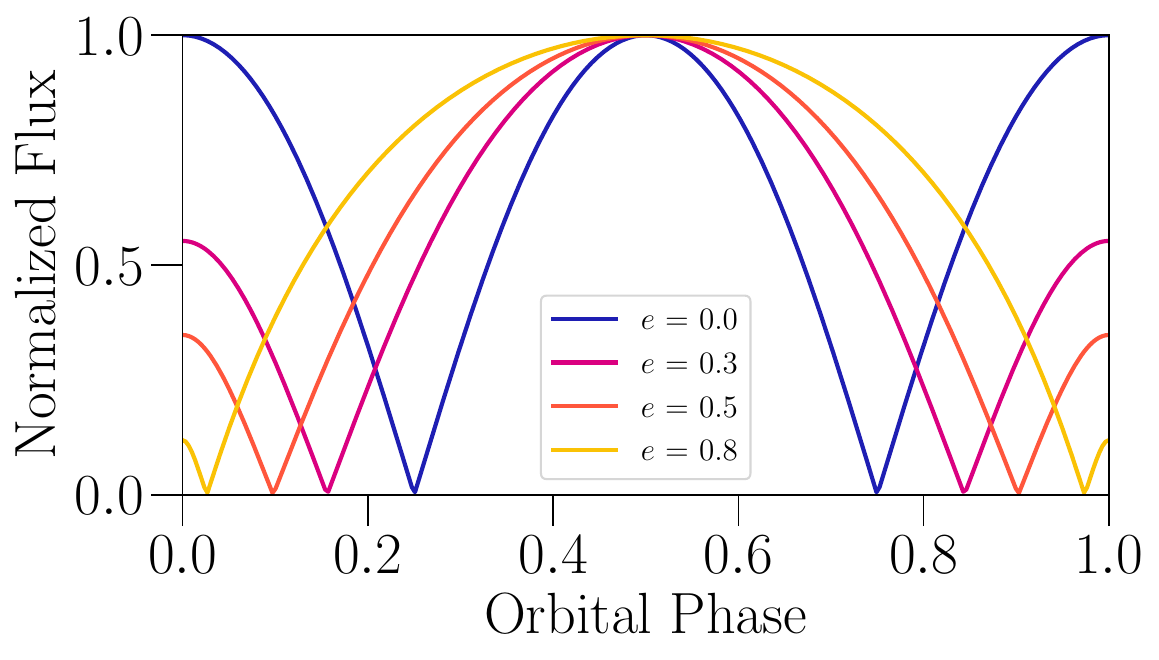}
    \caption{Synthetic light curves showing incoming flux to a polar-aligned disk as a function of binary orbital phase, plotted for various binary eccentricities.  All curves are normalized to their peak brightness value, which occurs at binary apocenter.}
    \label{fig:analytic_lightcurve}
\end{figure}

Figure \ref{fig:analytic_lightcurve} shows the total stellar flux on the disk for various initial binary eccentricities.  The light curve of the binary is split unevenly into two peaks corresponding to the two orbital arcs on either side of the disk, covering binary pericenter (phase=0) and apocenter (phase=0.5).  In the limit of a circular binary $(e=0)$, the flux shows regular periodicity at exactly half the binary orbital period.  The incoming flux drops to zero twice each binary orbit, corresponding to the two disk plane crossings.  As the binary eccentricity increases, both the periodicity and the amplitude of the flux become increasingly uneven.  The peak at binary apocenter grows in amplitude and duration, while the peak at binary pericenter shrinks.  For a binary with an eccentricity of $e=0.5$, the passage through pericenter takes about 20 percent of the total orbital period, while the arc through apocenter completes the other 80 percent.  Polar alignment is more likely to occur for disks around eccentric binaries, suggesting the flux curves observed from these disks are likely to be asymmetric along the orbital period.

\subsection{Disk flaring}
A more realistic depiction of the disk geometry includes the finite disk thickness and flared disk geometry.  A flared disk geometry gives all parts of the disk a direct line of sight to the center of the system, reducing the amount of self-shadowing.  Additionally, the surface normal $\mathbf{\hat{n}}$ is angled inward towards the center, increasing the amount of illumination received from the central stars.  The amount of flaring in a disk is typically characterized by a flaring index $\beta$, where the disk scale height $h/r \propto r^\beta$.  A positive value for $\beta$ indicates disk flaring, while $\beta = 0$ represents a disk with a constant scale height (no flaring).

To include flaring geometry in our analytic calculations, we parameterize the disk height using an initial disk scale height $h_0$ and a flaring index $\beta$, and set the vertical height of the disk as $h(r) = h_0 (r/r_0)^{\beta+1}$.  We also consider the changing angle of the disk surface, which is angled to the vertical axis by an amount $dh(r)/dr = (\beta + 1) h/r (r)$.  The third row of Figure \ref{fig:diskillumination}  shows the illumination profile for a disk with a scale height of $h_0 = 0.05$, $r_0 = a$, and flaring index $\beta = 0.25$, a structure similar to passively illuminated disks \citep{Chiang1997}.  Flaring disks have a similar flux pattern as a flat disk, but receive slightly more flux overall, causing a shallower falloff in flux farther out in the disk.  The flux does not fall to zero during disk plane crossings, as the flaring profile allows all parts of the disk to maintain a direct line of sight to the star.

\subsection{Inner rim}
The inner edge of a protoplanetary disk is the only portion of the disk that receives direct illumination at all times, even in the single star case when the star remains fixed in the disk plane.  The large amount of thermal energy received by this portion of the disk may change the geometry of the inner portions of the disk.  In some models, this region is thought to ``puff up'' the inner rim of the disk, increasing the vertical scale height and potentially casting a shadow on large portions of the outer disk \citep{Isella2005,Dong2015}.

In a polar disk scenario, the inner rim will still receive a large amount of stellar radiation, potentially leading to a puffed up inner rim.  The effect on the disk illumination is small when the stars are at apocenter, since light is coming from high above the disk, but the presence of an inner rim may be more important for intercepting starlight during binary pericenter and disk plane crossings, when the vertical displacement of the stars is low.

To investigate these effects, we model an inner puff in a similar fashion to the artificial rim used in \cite{Dong2015}, as a solid wall that blocks all incoming light.  We add an inner rim to our flared disk model of height $h_{\rm puff} = 1.5h_{\rm in}$, where $h_{\rm in}$ is the nominal height of the inner edge of the disk following the standard power-law profile.  When calculating the flux onto the disk, we check the vertical height of the rays as they cross the inner rim; incoming light rays with a vertical height less than the rim height $h_{\rm puff}$ are removed.  As stated in \cite{Dong2015}, this treatment is artificial and overestimates the effect of the rim on the illumination of the disk.  A more realistic rim will intercept only a portion of the stellar flux and instead cast a partial shadow on the outer disk.  We consider this model to represent the upper limit for the effect of shadowing from an inner rim.

In Figure \ref{fig:diskillumination}, the dashed lines in the third row highlight the regions of the disk which would be shadowed by a heightened inner rim.  The gray and white contours correspond to puff heights of $h_{\rm puff} = 1.5h_{\rm in}$ and $2.0h_{\rm in}$, respectively.  The effect is negligible when the stars are far from the disk plane, but large shadows are cast at times close to disk plane crossing; for $h_{\rm puff} = 2.0h_{\rm in}$, the shadow covers nearly the entire disk.  Including the effects of shadowing is important in determining an accurate amount of flux during disk crossing, since both stars are in the disk plane and will be partially shadowed.

\begin{figure}
    \centering
    \includegraphics[width=0.9\columnwidth]{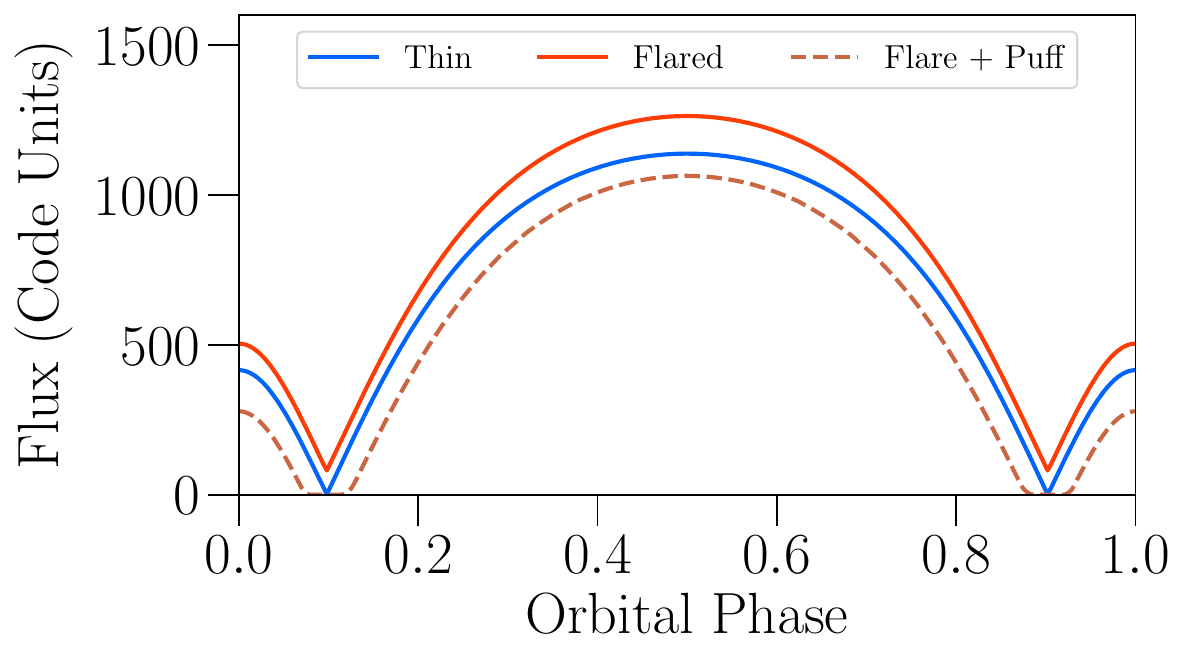}
    \caption{The effect of different disk models on the phase light curve for a disk around an $e=0.5$ central binary.  Unlike Figure \protect\ref{fig:analytic_lightcurve}, values are not normalized in order to highlight the differences in brightness between the models.  Flared disks appear slightly brighter than the razor-thin disk, while puffed disks show a longer minimum during binary pericenter.}
    \label{fig:lightcurvecomp}
\end{figure}

In order to highlight the effects of the different geometric models considered in this section, we plot the flux curves for each model in Figure \ref{fig:lightcurvecomp}.  The flux curves are created using a disk similar to HD 98800B $(r_{\rm in} = 2a, r_{\rm out} = 5a)$ but using a central binary of $e = 0.5$ to highlight the effect of the binary orbit.  We use an initial scale height of $(h/r)_0 = 0.05$ at $r=a$.  For the flaring and puffed disk models, we apply a flaring index of $\beta = 0.25$, and increase the scale height of the inner edge of the disk to $h_{\rm puff} = 2.0 h_{\rm in}$ for the puffed disk model.  Flared disks intercept more light than the razor-thin approximation, as the disk surface is more directly illuminated by the starlight.  For the puffed disk model, the disk receives less light overall and has a more pronounced minimum during binary pericenter, with the incoming flux dropping to zero for an extended period of time during plane crossings.  If the inner bump is high enough, it is possible for the flux to drop to zero completely for the entirety of binary pericenter, removing the secondary peak entirely.  This suggests that the stellar flux is intercepted entirely by the inner rim during the pericenter passage.  However, each model is quite similar, and may be difficult to differentiate when comparing with observations.

\section{Radiative transfer modeling}
\label{sec:radmc}

In the previous numerical analysis, there are several additional effects which we do not consider for the sake of simplicity in our calculations.  The combination of finite star size and disk thickness, positioning of the second star during pericenter, and scattering within the disk create a complex series of self-shadowing and secondary lighting effects along multiple different regions of the disk, the sum of which is difficult to calculate via analytic methods.  To try and capture many of these behaviors at once, we turn to a Monte Carlo radiative transfer code which can capture the total behavior of these effects all at once.

We perform radiative transfer modeling using the software \textsc{radmc3d} \citep{Dullemond2012}.  We modify the default \textsc{ppdisk} setup to generate a polar-aligned protoplanetary disk.  The simulation grid consists of a $64 \times 128 \times 256$ grid in spherical-polar coordinates $(r, \theta, \phi)$.  The simulation domain covers $[2\rm{au}, 10\rm{au}]$ in $r$, $[0, \pi]$ in $\theta$, and $[0, 2\pi]$ in $\phi$, with cells logarithmically spaced in $r$ and linearly spaced in $\theta$ and $\phi$.

The setup of our disk is made to mimic the circumbinary disk of HD 98800B.  We set the inner and outer radii of the disk at 2.5 and 5 au, respectively.  We consider two cases where the initial scale heights of $(h/r)_0 = 0.01$ and $0.05$ at $r=r_{\rm in}$.  The disk is initialized in the $xy$-plane of the simulation domain, following a surface density power-law profile of $\Sigma \propto r^{-1}$.  The dust in the disk is distributed between $0.1 \um$ and 10.0 mm, with a power-law exponent of $-3.5$.  We adopt a dust-to-gas mass ratio of $1 \times 10^{-2}$ solar masses, and assume the dust is perfectly mixed with the gas.  Note that by assuming perfect mixing of the dust, the dust and gas scale heights are identical.  For the dust opacities, we use the DSHARP opacities from \cite{Birnstiel2018}.  We ignore contributions from ice grains, as the relatively compact size of the disk keeps it within the ice line.

The central binary of HD 98800B is highly eccentric $(e =0.8)$, with an orbital separation of 1 au.  We treat the stars as two central light sources which move with time.  For the two components (Ba and Bb), we use stellar masses of $M_{\rm Ba} = 0.77M_\odot,\ M_{\rm Bb}=0.58M_\odot$, stellar radii of $R_{\rm Ba}=1.09R_\odot,\ R_{\rm Bb}=0.85R_\odot$, and  stellar temperatures of $T_{\rm Ba}=4200\rm{K},\ T_{\rm Bb}=4000\rm{K}$ \citep{Boden2005,Zuniga-Fernandez2021}.  We analytically solve for the orbits of the stars by solving Kepler's equation, generating 100 positions for each star equally spaced in time.

Similar to Section \ref{sec:lightcurves}, we consider three disk models in our radiative transfer simulations.  Our first model is a non-flaring disk, with $\beta = 0$ and a constant scale height throughout.  The second model considers a flaring disk with $\beta = 1/4$.  Our third model adds a puffed inner rim to the flared disk model, created by doubling the initial scale height in the range of $r$ = 2 au to 3 au.

We generate synthetic images of the disk using \textsc{radmc3d} across a full orbit of the binary in two steps, first by calculating the thermal structure of the disk with the given stellar positions, then ray tracing the image at a given wavelength.  This process creates images under the assumption that the gas and dust density field is static with time, and that the emission from the disk responds instantly to the changing radiation field.
We generate synthetic images at three different wavelengths:  $1.65 \rm{\mu m}$ (H-band), $0.8$mm (ALMA Band 7), and $1.3$mm (ALMA Band 6).  These wavelengths correspond to images taken at scattered light and millimeter/sub-millimeter wavelengths, and are similar to the observational data of HD 98800B taken in the past (see Appendix \ref{sec:appendix}).  We use $10^7$ photon packages to generate our synthetic images, with scattering of photons treated anisotropically using the Henyey-Greenstein function \citep{Henyey1941}.  We utilize the modified random walk method (MRW, \citealt{Fleck1984}) for photon transport to reduce computation time in optically thick regions.

\section{Results}
\label{sec:results}

\begin{figure}
    \centering
    \includegraphics[width=1.0\columnwidth]{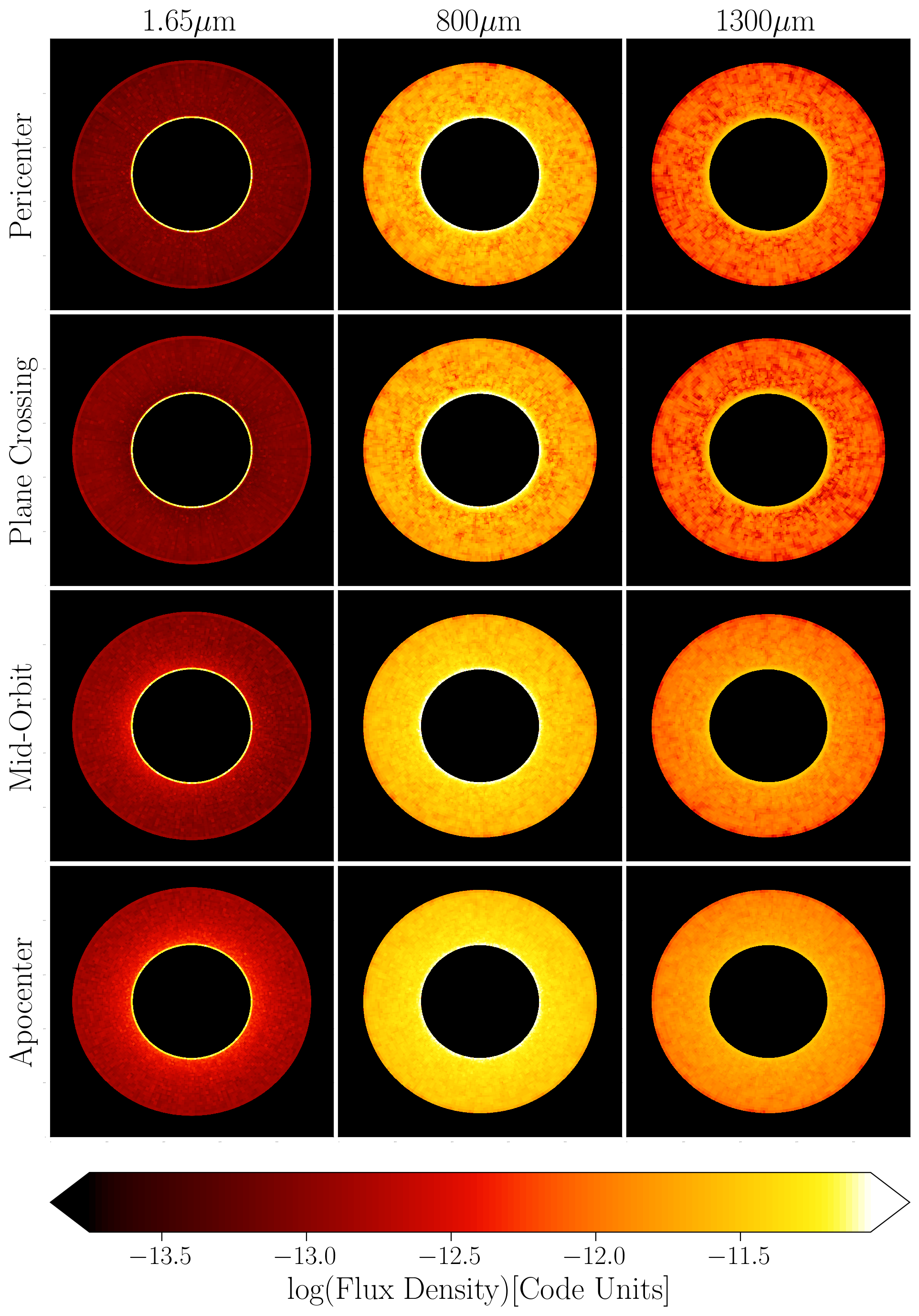}
    \caption{Synthetic radiative transfer images created from \textsc{radmc-3d}.  Each column shows the disk at a particular observing wavelength, while each row shows the disk at a particular point in the binary orbit.  From top to bottom, the images are shown at:  binary pericenter, during disk plane crossing, mid-orbit, the point of greatest horizontal displacement of the binary, and binary apocenter.  From left to right: observing wavelengths of $1.65\mu$m (H-band), $800\mu$m, and 1.3mm (ALMA Band 7).}
    \label{fig:images_radmc3d}
\end{figure}

Our radiative transfer simulations exhibit brightness changes across multiple wavelengths during the binary orbital period, in agreement with our numerical analysis in Section \ref{sec:lightcurves}.  In Figure \ref{fig:images_radmc3d}, we show the results of our radiative transfer simulations for a flared disk model.  Each row plots the flux density $F_\lambda$ of the disk at a particular wavelength, while each column shows the appearance of the disk when the binary is at different orbital phases (pericenter, disk plane crossing, mid-orbit, and apocenter).  The synthetic images show many similarities to the images produced in Figure \ref{fig:diskillumination}, showing considerable dimming of the disk face as the stars approach pericenter.  The secondary brightness peak at pericenter is not as visible in the radiative transfer images.  A small amount of azimuthal asymmetry is visible in the H-band images at mid-orbit (third row), when the stellar components are both horizontally and vertically offset from the center of the system.

\begin{figure}
    \centering
    \includegraphics[width=0.95\columnwidth]{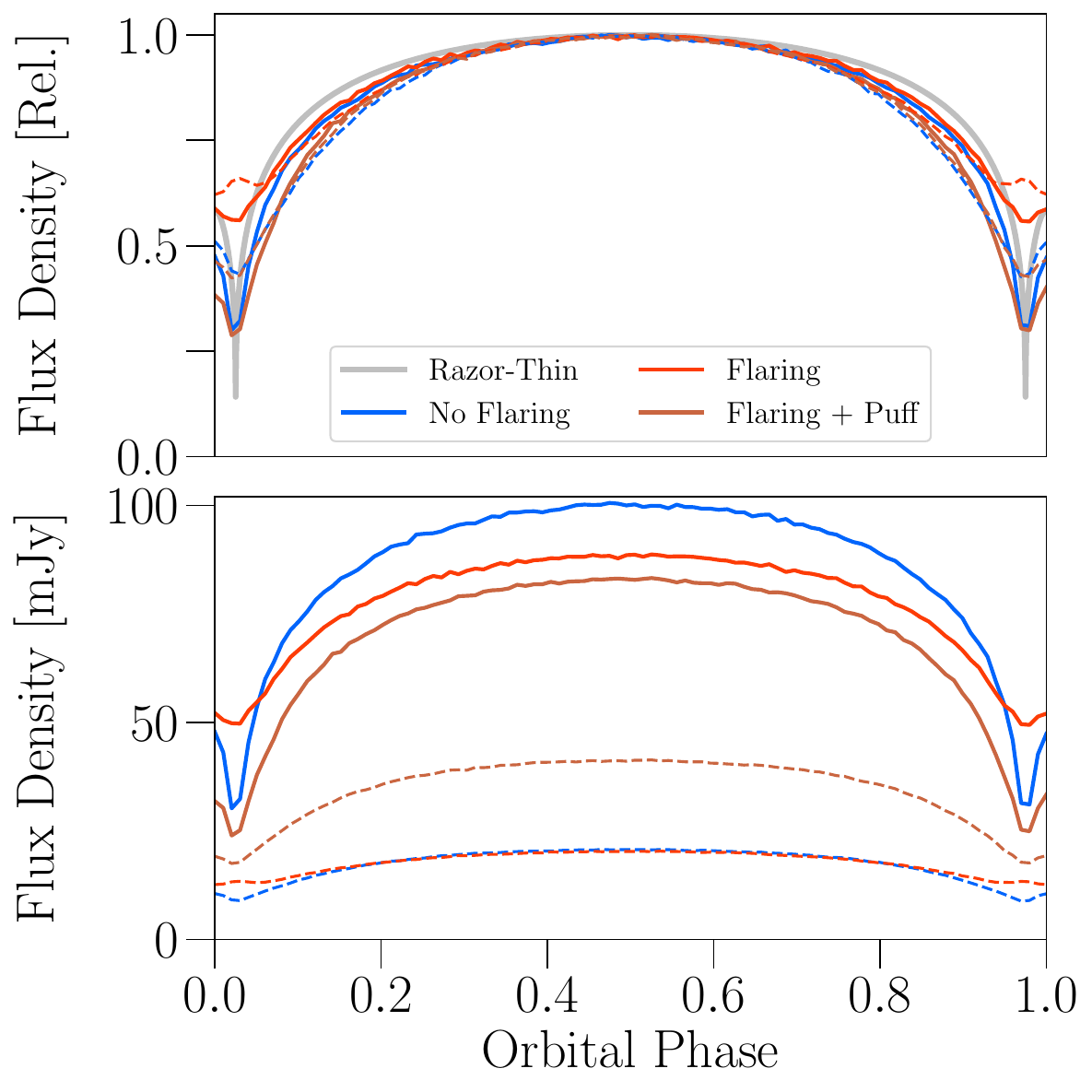}
    \caption{Comparison of 1.3mm and H-band lightcurves created from the \textsc{radmc-3d} radiative transfer models, with a disk scale height of $h/r=0.01$. In the top panel, curves normalized to their peak brightness at apocenter to emphasize the amount of variability.  For the top panel we also show the numerical flux calculated from the razor-thin disk model (Fig. \protect\ref{fig:analytic_lightcurve}), with an exponent of 1/4 to represent disk temperature and thermal emission at millimeter wavelengths.  In the bottom panel, we plot the absolute flux density assuming the disk is viewed face-on at a distance of $d=45$ pc, the distance of HD 98800B.  Solid and dashed curves indicate observations at $\lambda=1.3$mm and $1.65\um$ (H-band), respectively.}
    \label{fig:radmodelcomp}
\end{figure}

\begin{figure}
    \centering
    \includegraphics[width=0.95\columnwidth]{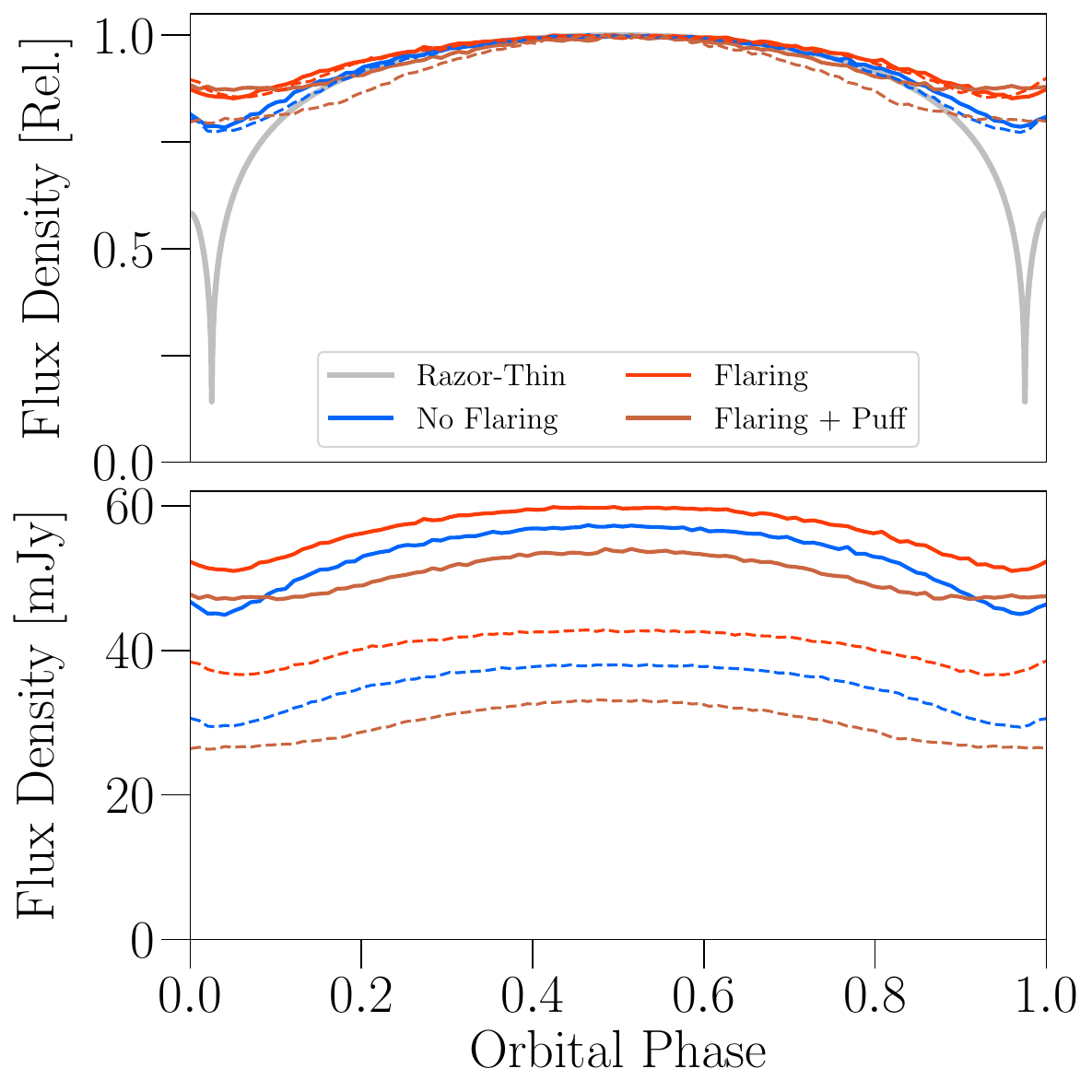}
    \caption{Same as Figure \protect\ref{fig:radmodelcomp}, but with $h/r=0.05$.}
    \label{fig:radmodelcomp_hr05}
\end{figure}

Figure \ref{fig:radmodelcomp} shows how our radiative transfer simulations are modified by different disk geometries for an inner disk scale height of $h/r=0.01$.  We plot synthetic lightcurves at both H-band and ALMA Band 7, showing our disk models with no flaring, flaring, and a flared disk with an inner puff.  In the top panel we show the normalized flux density relative to the peak at binary apocenter, while in the bottom panel we show the absolute flux density assuming the disk is viewed face-on at the distance of HD 98800B ($d=45$pc).  In gray we plot the analytic thermal flux from a razor-thin disk with an $e=0.8$ central binary, using the incident flux curve generated from Figure \ref{fig:analytic_lightcurve}.   Assuming the disk is optically thick, the disk should radiate as a blackbody, with the temperature of each part determined by the local radiation field.  Combined with our assumption of thermal equilibrium with the stellar radiation field, this temperature corresponds to the incident stellar flux on the disk through the Stefan-Boltzmann law, $F_{\rm in} \propto \sigma T^{1/4}$.

All three disk models create light curves which are similar to the razor-thin model, with the puffed disk model predicting a slightly lower relative flux density due to self-shadowing effects.  The largest difference in the various models appears during pericenter, where the dimming of the disk is strongly determined by the disk geometry.  Flaring disks capture more of the stellar flux and exhibit a low amount of variability, dimming by less than a factor of two, while the flat and puffed models exhibit strong self-shadowing effects, dropping to nearly four times their peak brightness at apocenter.  The pronounced minima in these models also allow for the secondary peak at pericenter to be observed.  For a thicker disk (Figure \ref{fig:radmodelcomp_hr05}), the variability is significantly reduced for all models, with the disk dimming to only 80 to 90 percent of its maximum brightness during plane crossings.  When comparing the absolute flux density from the models, it is notable that the light curves brighten considerably as the disk scale height decreases.  Our models show differing levels of brightening, with the non-flaring disk becoming brighter than the flaring model at small $h/r$.  We attribute this effect to the changing illumination angle.  In a standard model of a flaring disk (i.e. \citealt{Chiang1997}), stellar photons strike the disk at a shallow, ``grazing'' angle.  This shallow angle allows the photons to penetrate a significant distance into the disk before being scattered, transferring energy deeper into the lower layers.  When the star is at apocenter, its vertical displacement can be greater than the vertical height of the disk, and stellar photons shine more directly down into the surface.  Photons emitted here will encounter high-density material faster and are absorbed higher up in the disk, keeping the midplane at a cooler temperature and reducing the millimeter flux density.  This can also explain why the flared model is brighter than the non-flared model during binary pericenter; it is only when the binary is far above the disk that the assumption of a grazing angle breaks down.

\begin{figure}
    \centering
    \includegraphics[width=1.0\linewidth]{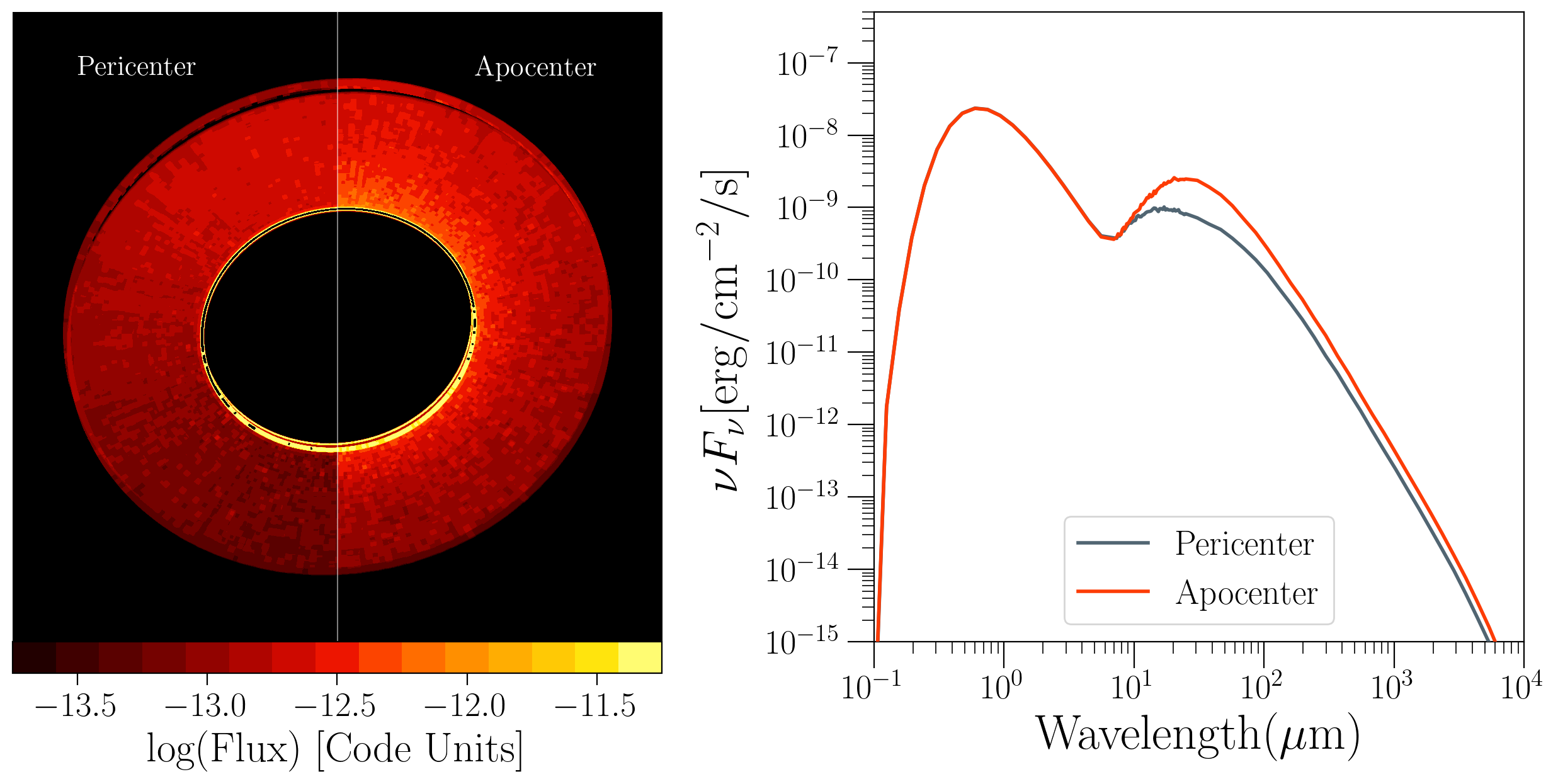}
    \caption{Comparison of a disk similar to HD 98800B in H-band emission at binary pericenter and apocenter. In this figure the disk is flaring, with a scale height of $h/r=0.01$, and has been rotated to match the on-sky appearance of the HD 98800B.  The left panel shows the emission across the surface of the disk (left:pericenter, right:apocenter), while the right panel shows the SED of the system (gray:pericenter, orange:apocenter).}
    \label{fig:sedcomp}
\end{figure}

In Figure \ref{fig:sedcomp} we compare the appearance of the disk at binary pericenter and apocenter by examining the synthetic images in H-band emission (left panel) as well as the SED of the system (right panel).  For this figure, we rotate the disk to match the orientation of HD 98800B as it appears on the sky $(i=154^\circ, \rm PA = 16^\circ)$.  In addition to the overall brightness increase, the scattered light images show more even illumination when the stars are farther out of the disk plane.  The observed changes are due to the vertical motion of the stars changing the scattering phase angle.  Near binary pericenter, the incoming starlight impacts the disk at shallow angles, emphasizing forward scattering and brightening the near side of the disk, as seen in many scattered light images of circumstellar disks \citep{McCabe2002,Benisty2023}. Close to apocenter, the increased height of the star increases the scattering angle on the near side of the disk, reducing the amount of forward scattering and causing the disk to appear closer to a smooth ring.  During binary apocenter, the peak of the SED, which sits at roughly $20 \um$, brightens by over a factor of 2, with similar increases in flux density present out into the millimeter wavelengths.

Our radiative transfer models suggest variability for both the HD 98800 system and other sources with polar-aligned disks, visible at multiple different observing wavelengths.  The geometry of the disk can modify the amount of brightness variations seen in the disk over the binary orbital period.  If the thermal response of the disk is fast, the variability may be visible at mm wavelengths.  In addition, smaller effects involving azimuthal asymmetries in scattered light may also be present.

\section{Discussion}
\label{sec:discussion}

\subsection{Using variable illumination to probe inclined disks}

Flux variability caused by the motion of the central stars out of the disk plane is most noticeable when considering polar-aligned disks, but it is also likely to occur in all disks with a central binary.  Even though we focus on HD 98800B in this paper, where the binary and disk planes are perpendicular, it may be possible to detect similar illumination patterns in other disks with a smaller mutual inclination to their central binaries.  A handful of well-resolved disks have been measured with a significant mutual inclination $\incmut$ between the disk and (outer) binary planes, including the GW Orionis circumtriple disk ($\incmut = 28.8^\circ$, \citealt{Rabago2024}), the circumbinary disk of HD 142527 ($\incmut = 50^\circ$, \citealt{Nowak2024}), and the circumtriple disk of GG Tau A ($\incmut \sim 30^\circ$, \citealt{Toci2024}).  In circumtriple systems, polar alignment much more difficult to maintain, and is only expected to be stable in a close-in radial region around 3-10 times the separation of the outer binary \citep{Ceppi2023,Lepp2023}.  Observations of other circumbinary disks may reveal this phenomenon, and may even help confirm the misaligned nature of the disks themselves. A degeneracy in the longitude of the ascending node as measured on the sky plane commonly leads to two differing values for $\incmut$, and examining the varying illumination on the disk surface may help constrain the true misalignment angle.  Observations taken close to disk plane crossing may provide constraints on the disk scale height and cooling time.

We note that the images and light curves created in our radiative transfer analysis (Section \ref{sec:radmc}) are created under the assumption that the thermal temperature of the disk at a particular moment corresponds to the radiation field of the stars at their current position, and that the density profile of the disk does not change during the binary orbit.  In other words, our models assume a disk with a quickly changing thermal structure but a slowly evolving density structure, providing a simplified scattering surface and thermal profile which can be easily examined.  Although a detailed calculation of the disk's density and thermal structure requires treatment of coupled, time-dependent radiation and hydrodynamic effects, we discuss the limitations of our models below.

For scattered light observations, the assumption of an instantaneous response from the disk may be reasonable as the observed image is just the light scattered off of the disk surface.  For observations at millimeter wavelengths, this assumption may be considered the limit of fast cooling rates in the disk.  If the rate of thermal cooling in the disk $t_{\rm cool}$ is comparable to the binary orbital period $\Tb$, then the temperature of the disk may relax considerably during pericenter passage, and the observed flux density at millimeter wavelengths may vary as in Figure \ref{fig:radmodelcomp}; on the other hand, slower cooling rates are more likely to keep the disk close to a steady temperature state and show less variability at mm wavelengths.  Figures \ref{fig:radmodelcomp} and \ref{fig:radmodelcomp_hr05} suggest that the amplitude of the flux density at mm wavelengths is highly dependent on the dust scale height, with $(h/r)_{\rm d} \lesssim 0.05$ required for amplitudes of around 10 percent.  Small dust scale heights in protoplanetary disks may be achievable with vertical dust settling, with theoretical works \citep{Youdin2007,Yang2021,Baronett2024} and recent observations \citep{Isella2016,Doi2021,Villenave2022} suggesting $h_{\rm dust} \sim 0.1 h_{\rm gas}$ in the dust rings of some disks $\left( h/r_{\rm gas} \approx 0.05-0.1 \right)$.  

Analytic estimates of $\tcool$ suggest fast cooling may occur in the inner regions of the disk.  We can estimate the cooling timescale of the disk by comparing the heat content of the disk and its luminosity (e.g. \citealt{Kratter2016}):

\begin{equation}
    \tcool = \frac{4}{9}\frac{1}{\gamma (\gamma-1)}\frac{\Sigma c_s^2}{\sigma T_{\rm eff}^4}.
    \label{eq:tcool_KL}
\end{equation}
Here, $\Sigma$ is the surface density of the disk, $c_s$ is the local sound speed, $T_{\rm eff}$ is the local effective temperature (which is equivalent to the surface temperature if the disk is optically thick), $\sigma$ is the Stefan-Boltzmann constant, and $\gamma$ is the adiabatic index.  This equation relates the total thermal energy of the disk and the energy outflow rate lost via radiation.  In the case of disk heating, which may happen as the stars rise out of the disk plane, there is instead a net energy influx, and we can replace the Stefan-Boltzmann component of the denominator with the incoming stellar flux, $L_*/4\pi r^2$.

\begin{figure}
    \centering
    \includegraphics[width=1.0\linewidth]{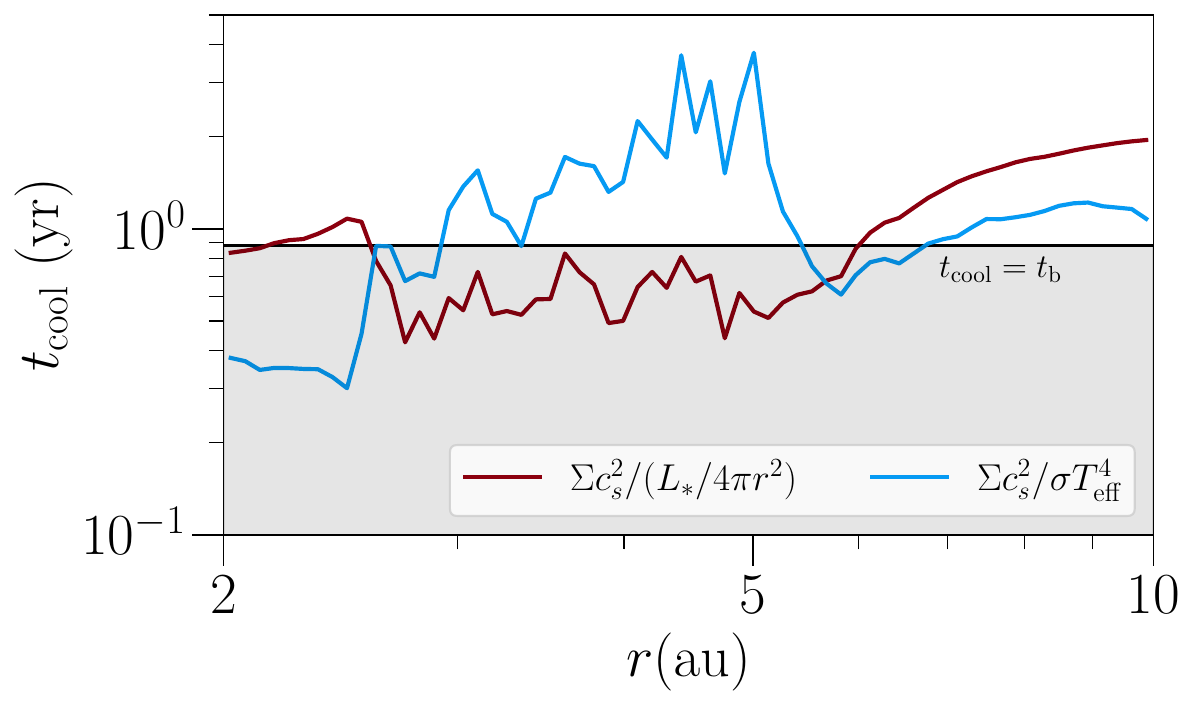}
    \caption{ Cooling timescale as measured from our radiative transfer models.  The blue and red curves represent timescales for radiative cooling and stellar heating, respectively.  The gray shaded region represents timescales which are shorter than the binary orbital period $\Tb$.  Note that the disk in our \textsc{radmc3d} models has a radial extent of 2.5 to 5au. }
    \label{fig:tcool}
\end{figure}

In Figure \ref{fig:tcool}, we plot the cooling and heating timescales for the disks used in the \textsc{radmc3d} models, using the disk surface density and temperature from the \textsc{radmc3d} radiative transfer calculation.  We note that this calculation is not entirely self-consistent, since the construction of the \textsc{radmc3d} models assumes equilibrium with the instantaneous stellar radiation field.  With these models, we find a cooling time of about 1 year at $r=2.5$au and 3 years at $r=5$au, very close to the orbital period of HD98800B.  Although this is a rough calculation, it suggests that the thermal timescales of the disk may be comparable to the binary orbital period in the inner regions of the disk, producing visible changes in the thermal flux.  We also note that opacity may play an important role in determining which regions of the disk are susceptible to thermal variation; a similar formulation of $\tcool$ by \cite{Zhang2020} includes effects for optically thick and thin limits, and predicts longer cooling times than the single opacity regimes at all radii.  If the emission is expected to be optically thick, which may be the case if the dust is settled into the disk midplane, then the optical depth is $\tau=\Sigma\kappa/2$, leading to shorter cooling timescales in the outer disk as the optical depth decreases and longer timescales in the inner disk as radiation is trapped at high optical depths.

Since these estimates predict $\tcool$ and $\Tb$ can be comparable, the inner disk may only partially cool during the binary orbit.  Thus, the variations in flux shown in Figures \ref{fig:radmodelcomp} and \ref{fig:radmodelcomp_hr05}, which assume instantaneous cooling, should be interpreted as an upper limit and that the actual observed fluctuations may be smaller.  The close competition between these two timescales strongly motivates the need for direct observations and long-term monitoring of misaligned disks to see if changes in the thermal flux of the disk are in fact visible and to better constrain the relevant cooling timescales.

The density profile of our disk models is assumed to be roughly static, as density fluctuations are expected to occur on the local orbital timescale $t_{\rm orb} = h/c_s$, which is longer than the binary orbital period $\Tb = \Omega_{\rm b}^{-1}$.  However, variations of the protoplanetary disk temperature with time have been found as a potential avenue for producing substructures in the outer disk.  Previous simulations have usually considered this effect in the context of shadowed lanes created by a misaligned inner disk \citep{Montesinos2016,Montesinos2018,Zhang2024,Ziampras2025,Zhu2025}, where material orbiting in and out of the shadow experiences sudden changes in temperature.  These shadowed lanes generate an asymmetric pressure gradient in the azimuthal direction, leading to the generation of spiral arms and vortices.  For a misaligned circumbinary disk, the motion of the central binary in and out of the disk plane may produce a similar effect, changing the irradiation flux across the entire disk.  In this case, the disk temperature may change globally instead of at a particular azimuthal location, but this effect may still have the potential to generate substructures.

\subsection{Observations of HD 98800}

\begin{figure}
    \centering
    \includegraphics[width=0.95\columnwidth]{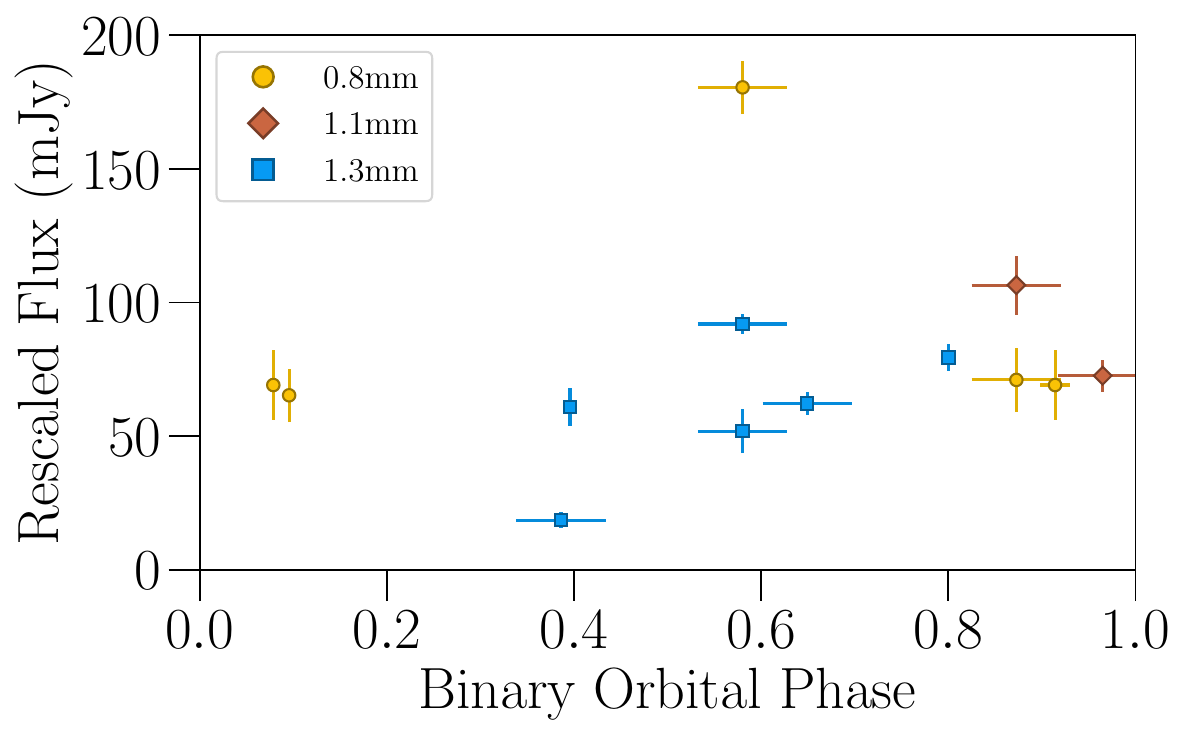}
    \caption{Archival observations of HD 98800B, with flux density rescaled using a spectral power-law index of $\alpha=2$.}
    \label{fig:observations}
\end{figure}

The HD 98800 system, and later HD 98800B, has long been known as a source presenting a strong infrared excess, with observations dating back over 35 years \citep{Walker1988}.  Observations of the system have been conducted across a large range of wavelengths, from optical \citep{Soderblom1998} to far-infrared \citep{Ribas2018}.  The system is known to be variable \citep{Soderblom1998}, though it is not clear if the variability is related to the binary orbital period.

Past observations of the HD98800B system are taken at various points in the orbital period, and thus may show variations in flux that is correlated with the orbital period.  To examine this idea, we search archival data for prior observations of HD 98800B at 0.8, 1.1 and 1.3mm wavelengths (for a full list of the observations considered here, see Appendix \ref{sec:appendix}). The date of observation, combined with an orbital solution of the BaBb binary, gives an orbital phase for each observation.  Assuming the flux density $F_\lambda$ at mm-wavelengths is optically thick, we rescale the observed $F_\lambda$ using a spectral index $\alpha$ using

\begin{equation}
    F_{\rm \lambda,rscl} = F_{\rm{\lambda,obs}} \left(\frac{\lambda}{1\rm mm} \right)^{-\alpha}
\end{equation}
where $F_{\rm{\lambda,obs}}$ is the flux density measured in the observation, $\lambda$ is the observing wavelength and $\alpha$ is the spectral index, which we choose as 2. Previous modeling of the HD 98800B disk has shown $\alpha=2$ to provide a good fit to the observed data out to cm wavelengths \citep{Ribas2018}, implying emission that is optically thick or from large dust grains.  We plot the rescaled flux density versus orbital period in Figure \ref{fig:observations}.  There is a large amount of scatter in the observations, and for the wavelength bands we consider there is not a clear trend of decreasing flux as the binary approaches pericenter.  However, the observations used to create Figure \ref{fig:observations} are a collection of observations using multiple different instruments across several years and different observing epochs, using different calibration techniques.  The mixture of different observing methods makes a proper comparison of the fluxes in this data set fairly challenging.

Future observations and long-term monitoring of the HD 98800B system across different orbital phases can provide increased insight to the illumination variability of the disk, as well as the response of the disk during the binary orbital period.  Improved coverage across the binary orbit, as well as better calibration and flux measurements from current instruments, will help constrain the intrinsic variability of the system.  The relatively short binary orbital period, $\Tb \sim 1\rm yr$, makes the HD 98800B system an easily accessible target for time monitoring.  Though scattered light observations are difficult due to the small size of the disk \citep{Rich2022,Cuello2025}, future observatories such as ELT may be able to resolve the disk and determine if any asymmetries are present.  For reference, Table \ref{tab:pericenter} lists the dates of future pericenter passages of the BaBb binary, as calculated from the two orbital solutions given in \cite{Zuniga-Fernandez2021} and \cite{Merle2024}.

From the orbital solution calculated in \cite{Zuniga-Fernandez2021}, the A binary of the HD 98800 system is expected to pass behind the circumbinary disk some time in the near future, with more recent modeling of the disk by \cite{Faruqi2025} suggesting the occultation event may have already begun.  This occultation event is expected to last several years, into the 2030s.  In combination with the analysis presented above, the HD 98800B disk is positioned as a target of particular interest for determining several fundamental disk properties.  Optical/NIR monitoring of the system in the coming years will provide a wealth of information into the structure of HD 98800B, and how the surrounding binary stars influence its evolution.  As described in \cite{Faruqi2025}, observations of how the A binary disappears behind the disk and reappears within the cavity can provide constrains on the optical thickness of the disk material, and may give hints into the presence of any perturbation-induced substructures.  Simultaneously, observing changes in the disk flux due to the B binary can provide constraints on the cooling rate of the disk material and the vertical structure of the disk.

\begin{table}
    \centering
    \caption{Recent and upcoming pericenter passages of the HD 98800B binary, using two different orbital solutions. }
    \begin{tabular*}{0.95\columnwidth}{l l}
         \cite{Zuniga-Fernandez2021} & \cite{Merle2024} \\
         \hline
         2024-01-28 & 2024-02-16\\
         2024-12-07 & 2024-12-27\\
         2025-10-17 & 2025-11-07\\
         2026-08-28 & 2026-09-17\\
         2027-07-08 & 2027-07-29\\
    \end{tabular*}
    \label{tab:pericenter}
\end{table}

\section{Conclusions}
\label{sec:conclusions}

Polar-aligned circumbinary disks are systems where the disk and central binary stars inhabit highly misaligned orbits.  This unique geometrical configuration allows the stars to rise vertically above and below the disk plane, potentially causing large amounts of variability in the disk illumination and temperature structure.  We show this variability is directly correlated with the motion of the binary using both numerical analysis and radiative transfer modeling, and produces a periodically varying lightcurve which peaks twice at binary apocenter and pericenter.  The geometry of the disk strongly affects the amount of stellar flux received by the face of the disk, which can modify the amount of variation in the observed lightcurve and change the length of the minimum at pericenter.  In scattered light observations, additional observational effects may be visible due to changes in the scattering angle as the stars move vertically, rendering the scattered light more asymmetric at pericenter.

Our radiative transfer simulations suggest the presence of many potential observable signatures which may be identified through long-term monitoring of misaligned disks.  Observing inclined disks at multiple epochs may reveal new insight into the geometry and internal structure of the disk, deepening our understanding of how the disks operate in other systems.  Flux variability at millimeter wavelengths can reveal the cooling rate of the disk during the binary orbit, providing insight into the internal thermal structure of the disk.  The circumbinary disk around HD 98800B is of particular interest, due to the presence of a nearly polar disk around a highly eccentric binary; the unique configuration of this system may allow future observations of this system to constrain multiple disk parameters at the same time.

\begin{acknowledgements}
We thank the anonymous referee for their suggestions which have improved the overall quality of the manuscript.  I.R. thanks Rebecca Nealon for helpful discussions.  I.R. is funded by PRIN-MUR 20228JPA3A and by the European Union Next Generation EU, CUP: G53D23000870006. Views and opinions expressed are however those of the author(s) only and do not necessarily reflect those of the European Union or the European Research Council. Neither the European Union nor the granting authority can be held responsible for them.
\end{acknowledgements}




\bibpunct{(}{)}{;}{a}{}{,} 
\bibliographystyle{aa}
\bibliography{references} 



\begin{appendix}

\section{Past observations of HD 98800B}
\label{sec:appendix}

In Table \ref{tab:observationlist} we list the references for specific observations, including the observation date(s), wavelength band, total flux observed from the system, as well as the orbital phase derived using the orbital solution from \cite{Zuniga-Fernandez2021} \footnote{Note that the orbital solution from \cite{Merle2024} uses a slightly later starting epoch and a longer derived orbital period.  The cumulative effect of these differences is that the two solutions differ in the binary orbital phase by approximately 20 days, with the difference increasing by roughly half a day each year.  Using either orbital solution does not change the overall results.}.  For cases where the observation of the system is spread across multiple consecutive days, we use the midpoint of the observing campaign as the orbital phase and the beginning and end dates as the relative error bars.  In some cases, such as \cite{Jensen1996}, observations are listed as a sum of multiple campaigns throughout the year; for these cases, we list separate entries with different orbital phases for each observing session.

\begin{table}
    \centering
    \caption{Archival observations of the HD 98800B disk.  With the exception of \protect\cite{Kennedy2019}, all observations listed here are unresolved.}
    \begin{tabular*}{0.75\textwidth}{l r r r r}
         Reference & Observation Date(s) & Orbital Phase & $\lambda_{\rm obs}$ (mm) & Flux (mJy)  \\
         \hline
         \hline
         \cite{Rucinski1993}    & 25 Apr 1993   & 0.095    & 0.8   & $102 \pm 10$ \\
         \cite{Sylvester1994}   & Feb, Aug 1992 & 0.873     & 0.8   & $111 \pm 12$\\
         \cite{Walker1995}      & Jun 1994      & 0.580     & 0.8   & $282 \pm 10$\\
         \cite{Jensen1996}      & 24-27 Feb 1992& 0.914    & 0.8   & $108 \pm 13$\\
         \cite{Jensen1996}      & 26 Feb - 1 Mar 1993& 0.07    & 0.8   & $108 \pm 13$\\
         \hline
        \cite{Sylvester1996}    & Feb 1992      & 0.873       & 1.1    & $88 \pm 11$\\
        \cite{Sylvester1996}    & Mar 1992      & 0.965       & 1.1    & $60 \pm 6$\\
         \hline
         \cite{Stern1994}       & 17-18 Apr 1994& 0.395     & 1.3   & $36 \pm 7$\\
         \cite{Walker1995}      & Jun 1994      & 0.580     & 1.3   & $54.4 \pm 3.61$\\
         \cite{Sylvester1996}   & Apr 1994      & 0.386     & 1.3   & $11 \pm 3$\\
         \cite{Kennedy2019}     & 15, 19 Nov 2017& 0.800   & 1.3   & $47 \pm 5$
    \end{tabular*}
    \label{tab:observationlist}
\end{table}

\end{appendix}

\end{document}